\documentclass[
aps,
prd,
reprint,
showpacs,
superscriptaddress,
longbibliography,
floatfix,
nobalancelastpage,
nofootinbib
]{revtex4-2}

\usepackage[bbgreekl]{mathbbol}
\usepackage{amsfonts}
\DeclareSymbolFontAlphabet{\mathbb}{AMSb}
\DeclareSymbolFontAlphabet{\mathbbl}{bbold}

\usepackage{linenoAMS}
\usepackage{amsmath}
\usepackage{upgreek}
\usepackage{siunitx}
\usepackage{subfigure}

\usepackage[colorlinks=true, citecolor=RoyalBlue,
linkcolor=BrickRed, urlcolor=ForestGreen]{hyperref}
\usepackage[all]{hypcap}

\usepackage{enumitem}
\usepackage{autofigs9}
\usepackage{braket}
\usepackage{comment}
\usepackage{mathtools}

\graphicspath{{Figures/}}
\usepackage[usenames, dvipsnames, svgnames, table]{xcolor}

\usepackage{xfrac}
\usepackage{ulem}
\usepackage{cancel}
\usepackage{graphicx}

\usepackage[capitalise]{cleveref}
\usepackage{makecell}

\begin{document}
\title{High-efficiency Electro-Optic Lens for radio frequency beam wavefront modulation for mode mismatch sensing}

\author{Liu Tao}
\email{liu.tao@ligo.org}
\affiliation{University of Florida, 2001 Museum Road, Gainesville, Florida 32611, USA}
\author{Mauricio Diaz-Ortiz}
\affiliation{Donders Institute, Department of Neurophysics, Radboud University, Houtlaan4, 6525 XZ, Nijmegen, The Netherlands}
\author{Paul Fulda}
\affiliation{University of Florida, 2001 Museum Road, Gainesville, Florida 32611, USA}

\date{\today}

\begin{abstract}
Active mode mismatch sensing and control can facilitate optimal coupling in optical cavity experiments such as interferometric gravitational wave detectors. In this paper, we demonstrate a radio-frequency (RF) beam wavefront curvature modulation-based mode mismatch sensing scheme inspired by the previously proposed RF beam jitter alignment sensing scheme. The proposed mode mismatch sensing scheme uses an electro-optic lens (EOL) device that is designed to provide the required beam wavefront curvature actuation, as well as a mode converting telescope that rephases the RF second-order modes and generates a non-vanishing mode mismatch sensing signal. We carefully investigate the total second-order mode generation from the wavefront actuation both analytically and numerically, taking the effects of Gaussian beam size evolution and the second-order mode phase mismatch cancellation into consideration. We demonstrate the second-order mode generation as a function of the incident beam waist size and the electro-optic crystal size, which along with a ``trade-off'' consideration of the beam size at the edges of the crystal and the clipping loss, provides us with guidance for designing the beam profile that interacts with the crystal to improve the EOL modulation efficiency.

\end{abstract}

\maketitle
\section{Introduction}
Precise locking and control of perfect alignment and mode-matching states are essential for maintaining efficient power coupling of laser beams to spherical optical resonators in many high-precision optical cavity experiments, such as in searches for particles beyond the standard model~\cite{DIAZORTIZ2022100968} and precision measurements of quantum gravity phenomena at the Planck scale~\cite{Chou_2017, Vermeulen_2021}. This requires, for instance, that the laser beam wavefront curvature must match the curvature of the cavity mirrors at their positions on the optical axis to achieve optimal resonance in the optical cavity. Imperfect matching of the eigenmode of an optical cavity and an injected laser beam mode leads to degradations in optical resonance and power coupling for the laser beam~\cite{Bayer-Helms:84, Jones:20, Tao:21}.

High-finesse optical cavities are also widely used in applications related to gravitational wave detection. Monitoring and correcting the mode-matching state is becoming increasingly important for advanced Gravitational Wave (GW) detectors such as Advanced LIGO and Advanced Virgo~\cite{aLIGO, AdVirgo}, which are mostly composed of multiple inter-coupled suspended cavities, to yield the optimal sensitivity. For instance, it was demonstrated that the coupling of injected squeezed light in advanced GW detectors to the main interferometer can be extremely sensitive to mode mismatch losses~\cite{PhysRevX.13.041021, Barsotti_2019, McCuller2021, kunsHOMsqueeze}. In addition, the power loss scattered into the second-order modes
by a mode mismatch between the laser beam and the output mode cleaner cavity in aLIGO also directly corresponds to a loss of GW signal and an increased shot noise level at the GW readout. One should ideally monitor and limit the amount of mode mismatch to prevent the degradation of squeezing performance and of the sensitivity of GW detectors.

Several hardware and schemes for sensing and correcting mode mismatches have been proposed. For instance, radio-frequency (RF) quadrant photodiodes or bullseye photodiodes have been proposed and demonstrated for sensing and controlling the beam wavefront distortion, with the use of multiple sensors and Gouy phase telescopes for the complete mode mismatch degrees of freedom (DoFs), such as the beam waist size and waist position mismatch~\cite{PhysRevD.100.102001, Mueller:00}. An alternative novel method through the generation of RF higher-order mode sidebands has also been proposed for determining and correcting the alignment and mode matching states of the laser mode coupling to the eigenmode of optical cavities~\cite{Fulda:17, Ciobanu_2020}. These RF beam modulation-based sensing schemes have potential advantages over the traditional standard schemes by greatly simplifying the experiment with the use of a single-element photodiode for the complete sensing of all DoFs and do not require any additional Gouy phase telescopes. 

In this paper, we extend the RF beam jitter alignment sensing scheme proposed by P. Fulda et al.~\cite{Fulda:17}, to characterize and correct mode mismatches between laser beam mode and the eigenmodes of optical cavities. This QPD-free mode mismatch sensing scheme uses an electro-optic lens (EOL) device as a fast wavefront curvature actuator that provides the required RF beam wavefront curvature modulation~\cite{Fulda:17, master_thesis, PhysRevD.108.062001, Goodwin-Jones:24}. Second-order spatial mode RF sidebands are generated at \textit{twice} the mode separation frequency of an optical cavity. This guarantees the co-resonance in the cavity of one of the second-order mode sidebands with the carrier light in the fundamental mode. Complete RF lens (RFL) mode mismatch sensing signals can theoretically be obtained through the detection of the beat signal between the second-order mode RF sidebands from the wavefront modulation with the second-order mode carrier field from the static mode mismatch with only a single-element photodiode for both mode mismatch DoFs, at orthogonal demodulation phases, as described in our previous work~\cite{PhysRevD.108.062001}.

In this paper, we propose a design and demonstrate by simulation our implementation of the electro-optic lens device, by sandwiching an electro-optic crystal between three pairs of alternating polarity electrodes. We will calculate the change in the refractive index of the crystal near the central axial region due to the
Pockels electro-optic effect, and thus the accumulated phase map for the input Gaussian beam coupling, and show that it can be well approximated by a hyperbolic paraboloid surface. The interaction of the Gaussian beam with the hyperbolic paraboloidal phase map is characterized in terms of scattering in the Hermite-Gaussian (HG) mode basis through numerical mode decomposition technique. The phase of the second-order scattered modes $\mathrm{HG}_{2,0}$ and $\mathrm{HG}_{0,2}$ differ by $\pi$ due to the opposite signs for the focal length from the phase profile in the two principal transverse directions, which leads to vanishing RFL mode mismatch sensing signals. The hyperbolic paraboloid phase profile, and thus the phase difference between the second-order modes, can however be corrected by passing the beam through a subsequent astigmatic mode converter. This rephases the second-order modes in the RF wavefront modulation sidebands, producing the desired second-order Laguerre-Gauss $\mathrm{LG_{1,0}}$ mode for the RFL sensing.

The strength of the RFL mode mismatch sensing signals depends on the effective modulation depth, namely the amount of second-order modes in the RF beam wavefront curvature modulation sidebands as the Gaussian input beam interacts with the electro-optic crystal inside the EOL. On the other hand, the size and the wavefront curvature of the input beam vary as it propagates inside the crystal. This changes the amplitude and the phase of the RF modulation second-order modes. The phase mismatch between the second-order modes generated at different locations in the crystal leads to cancellation and a reduction in the total RF second-order mode generation. This cancellation of the RF second-order modes due to the phase mismatch is carefully calculated analytically in this paper. The result is verified with corresponding numerical results by treating the input Gaussian beam amplitude profile and the paraboloidal phase maps as two-dimensional arrays. This more careful and precise treatment gives us a significant difference in the estimation of the total RF second-order mode generation compared to the previous simplified approach by treating the entire interaction with the phase profile at a single location of the beam waist. It captures the effect of Gaussian beam size evolution and the phase mismatch cancellation and better characterizes the RF lens modulation efficiency. The total second-order mode generation as a function of the waist size of the input beam and the size of the electro-optic crystal, through a ``trade-off'' consideration with the beam size and consequently the clipping loss at the edges of the crystal, is demonstrated as a quantitative guidance for the design choice of the beam profile.

This paper is structured as follows: We start with our design and implementation of the electro-optic lens device for the proposed RF lens mode mismatch sensing scheme in Section~\ref{sec-1}. We then report a careful and precise calculation of the total second-order mode generation in Section~\ref{sec-2}, where the effects of beam size evolution and phase mismatch cancellation are demonstrated both analytically and numerically. In Section~\ref{sec-3}, we discuss the result and how it can guide us in designing the incident beam profile to achieve better RF second-order mode generation. We report conclusions and discussions for future work lastly in Section~\ref{sec-4}.

\section{Electro-Optic Lens Design}
\label{sec-1}
\begin{figure}[htbp]
    \centering
    \includegraphics[width=1\linewidth]{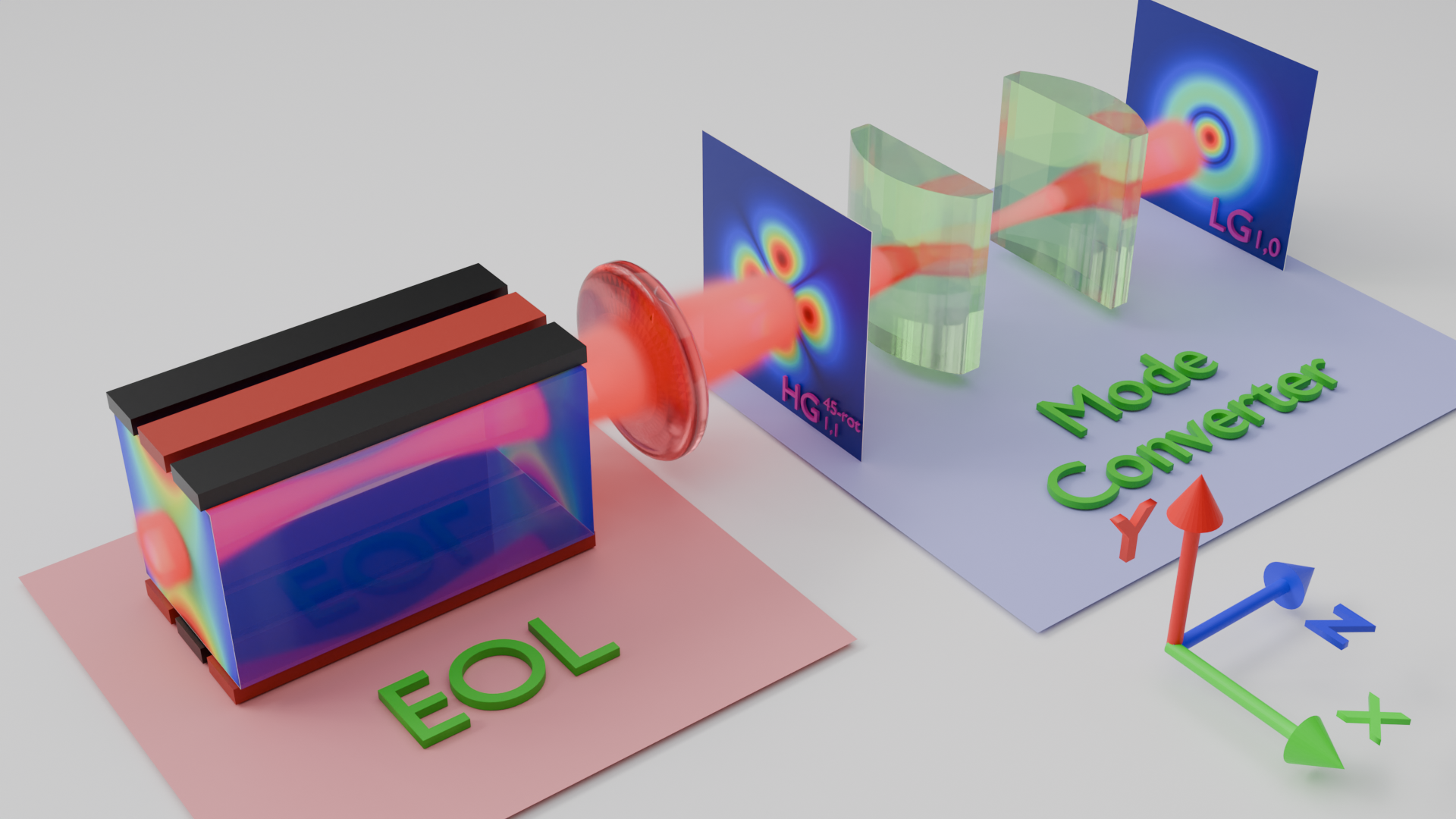}
    \caption{Illustration of RF lens modulation mode mismatch sensing scheme. The EOL design consists of an electro-optic crystal sandwiched between three pairs of alternating polarity electrodes (red for positive and black for negative). The input Gaussian beam picks up a hyperbolic paraboloidal phase profile as it passes through the electro-optic crystal, which generates the second-order modes $\mathrm{HG}_{2,0}$ and $\mathrm{HG}_{0,2}$ in anti-phase, characterized by the 45$^\circ$ rotated $\mathrm{HG}_{1,1}$ mode. The converted $\mathrm{HG}_{1,1}^{45^\circ \, \text{rot}}$ mode beam is reshaped by a mode-matching lens (red) and is focused to a beam waist at the center of the two cylindrical lenses that form the mode converter telescope (light green), with the focal length and separation of the lenses determined by the new waist size. It rephases the pair of second-order modes and converts them to the desired $\mathrm{LG}_{1,0}$ mode for the RFL sensing. }
    \label{fig-EOL_design}
\end{figure}

The proposed radio-frequency lens modulation mode mismatch sensing scheme implements an electro-optic lens device that can provide the required beam wavefront modulation. The electro-optic lens device can be designed based on three pairs of alternating polarity electrodes placed around a cuboid-shaped electro-optic crystal, such as the lithium niobate (LiNbO3) crystal. As illustrated in Figure~\ref{fig-EOL_design}, the positive electrodes are shown in red and the negative electrodes are shown in black. With a given voltage applied to the electrodes, it produces an electric field inside the crystal, causing a linear variation in the refractive index of the crystal, due to the Pockels electro-optic effect 
\begin{equation}
    \Delta n\left(E_{y}\right)=\frac{1}{2} n_{e}^{3} r_{33} E_{y}
    \label{equ-deltan}
\end{equation}
where $n_{e}$ is the extraordinary refractive index, and $r_{33}$ is the electro-optic coefficient of the crystal, with the coordinates defined in Figure~\ref{fig-EOL_design}. If we ignore the edge effect and assume a uniform electric field $E_{y}$ throughout the beam propagation direction $z$, we have $E_{y} = E_{y}(x, y)$. The input Gaussian beam picks up a total phase distortion to the wavefront as it propagates through the crystal, characterized by the following phase map
\begin{equation}
\Delta \phi = kL_{z}\Delta n\left(E_{y}\right)
   \label{equ-deltaphi}
\end{equation}
where k is the wavenumber, $L_{z}$ is the length of the crystal. The induced phase profile is proportional to the electric field distribution due to the linear electro-optic effect.

\begin{table}[htbp]
\centering
\caption{The parameters of the cuboid-shaped electro-optic crystal used in the EOL device.}
\begin{tabular}{c|c|c|c}
\hline \hline 
Parameter & Symbol&Value&Unit\\
\hline
\makecell{Extraordinary \\ index of refraction } & $n_{e}$ & 2.156 & - \\ \hline
Electro-optic coefficient & $r_{33}$ & 31 & pm/V \\ \hline
Physical Size & $L_{x}\times L_{y} \times L_{z}$ & $2\times2\times20$ & $\text{mm}^3$ \\ \hline
Electrode Size in x& $d_{x}$ &  0.653 & $\text{mm}$ \\
\hline
\hline
\end{tabular}
\label{tab-parameters}
\end{table}

\begin{figure}[htbp]
    \centering
    \includegraphics[width=1\linewidth]{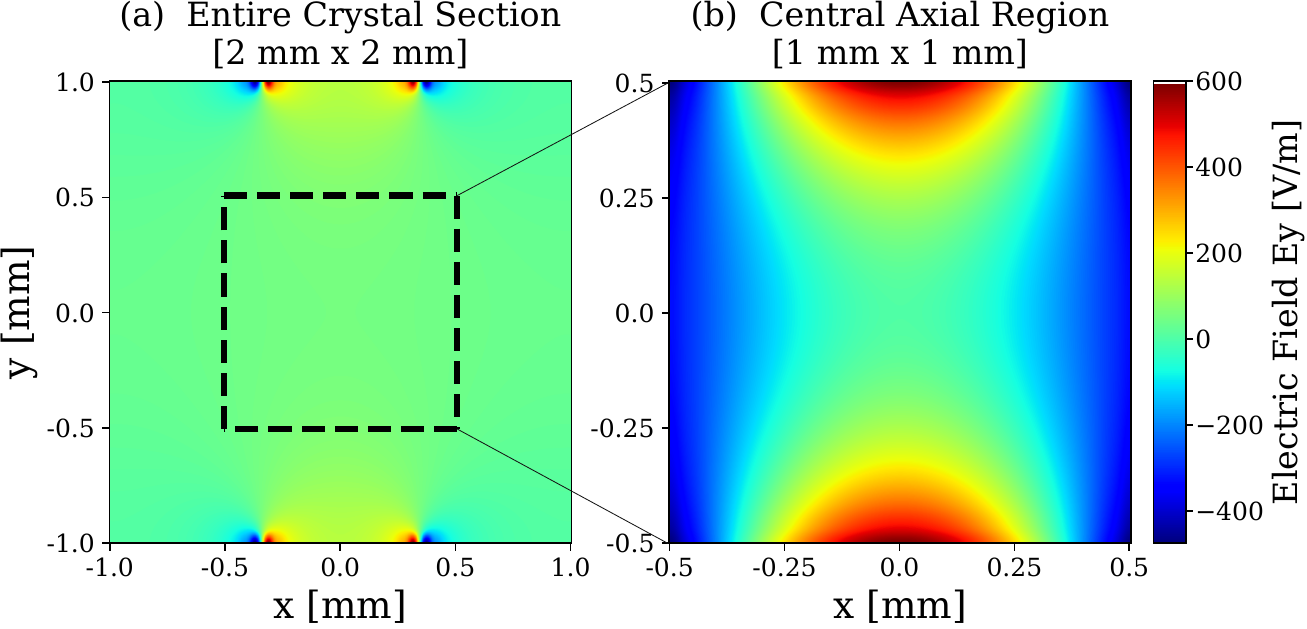}
    \caption{The numerical result on the electric field distribution inside the electro-optic crystal generated by applying $\pm 1$\,V to the electrodes obtained from finite element analysis simulation. Left: the entire cross section region of the crystal $2\,\mathrm{mm} \times 2\,\mathrm{mm}$; Right: the $1\,\mathrm{mm} \times 1\,\mathrm{mm}$ central axial region. The relevant properties of the electro-optic crystal used in the numerical modeling are listed in Table~\ref{tab-parameters}.}
    \label{fig-Ey_slice}
\end{figure}

With the boundary conditions from the electric potentials of the electrodes set to $\pm 1$ V, the electric field inside the crystal is solved numerically through finite element methods. Table~\ref{tab-parameters} lists the parameters of the electro-optic crystal used in solving the electric field distribution and the resulting phase profile. Figure~\ref{fig-Ey_slice} shows our numerical result for the electric field $E_{y}$ distribution inside the electro-optic crystal as a function of the transverse coordinates $x$ and $y$ obtained from finite element analysis modeling with the assumed crystal geometry and pairs of alternating polarity electrodes at the boundary. The left panel shows the numerical solution of the electric field distribution in the entire cross section region of the crystal, which is $2\, \mathrm{mm} \times 2\, \mathrm{mm}$ wide. The locations of the electrodes can be seen at the top and bottom, where the electric field is the strongest. The right panel shows the electric field distribution at the central $1\, \mathrm{mm} \times 1\, \mathrm{mm}$ region of the transverse direction, as illustrated by the black dashed square on the left. The electric field near the central axial region can be well approximated by a hyperbolic paraboloid surface. 

This electric field distribution produces a change in the index of refraction and consequently the extra phase accumulation profile that is proportional to $E_{y}$ according to Equation~\ref{equ-deltan} and Equation~\ref{equ-deltaphi}. The hyperbolic paraboloidal phase profile exerts an effect on the transmitted beam as a lens with equal magnitude but opposite sign focal lengths in the two principal transverse axes. In terms of HG mode scattering, sinusoidal modulation of the voltage applied to the electrodes produces pairs of RF wavefront modulation sidebands in both the second-order $\mathrm{HG}_{2,0}$ and $\mathrm{HG}_{0,2}$ modes, but in anti-phase with each other, due to the opposite sign in the focal length from the phase profile. 

To get a rough estimate of the amount of second-order mode generation, the total hyperbolic paraboloidal phase profile through the interaction with the entire crystal is applied to an input Gaussian beam at the waist location. The more precise and careful treatment is described in the next section. The phase profile and the beam amplitude distribution are represented in terms of two-dimensional arrays, and the mode decomposition is performed subsequently. For the beam profile that interacts with the electro-optic crystal, we start with a default design where the beam waist is placed at the center of the crystal. The beam sizes at the front and the back faces of the crystal are therefore equal due to symmetry. 

To achieve minimum clipping losses at the two ends of the crystal, one could be tempted to keep the size of the beam at the front and back faces of the crystal as small as possible; We will start with this particular configuration as our baseline nominal case and later show through detailed numerical and analytical calculations that this however is not necessarily the best choice. One can in principle achieve better RF beam wavefront modulation efficiencies through different beam profile configurations at the expense of a negligible increase in beam size at the faces of the crystal, and consequent clipping losses. The size of the input Gaussian beam at the edges of the crystal in general reads
\begin{equation}
\begin{aligned}
w(z) &= w_{0} \sqrt{1 + \left(\frac{z}{z_{R}}\right)^2} \\
&= \sqrt{w_{0}^2 + \frac{z^2 \lambda^2}{\pi^2 w_{0}^2}}
\end{aligned}
\end{equation}
where $z_{R}=\frac{\pi w_{0}^2}{\lambda}$ is the Rayleigh length. Thus the beam size at the crystal faces ($z = \pm\mathrm{L_z/2}$) reaches the global minimum when
\begin{equation}
 z_{R} = \frac{\pi w_{0}^2}{\lambda} = \frac{\mathrm{L_z}}{2} 
\end{equation}
Namely when the waist size is set to be $w_{m} = \sqrt{\frac{\mathrm{L_z}\lambda}{2\pi}}$, the front and end faces of the crystal are placed exactly at one Rayleigh length before and after the beam waist. For example, for a crystal that is 2 cm along the beam propagation direction, and for a wavelength of 1064 nm, the waist size that corresponds to the smallest beam size at the crystal edges is
\begin{equation}
w_{m} = \sqrt{\frac{\mathrm{L_z}\lambda}{2\pi}} \overset{\mathrm{L_z} = 2 \mathrm{cm}}{=\joinrel=} 58.2\, \mu m
\end{equation}
while the beam size at the crystal edges in this nominal case is $\sqrt{2}\cdot w_{m} = 82.3 \, \mu m$, which is the smallest beam size for a given crystal length of 2 cm.

\begin{table}[htbp]
\centering
\caption{The mode content by decomposing the input Gaussian beam applied with the hyperbolic paraboloid phase profile that is proportional to the electric field distribution in Figure~\ref{fig-Ey_slice} to the beam waist. The mode contents of the Hermite-Gauss modes are ranked by their amplitudes.}

\begin{tabular}{c|c|c|c|c}
\hline \hline 
HG Mode & $\mathrm{HG}_{0,0}$& $\mathrm{HG}_{0,2}$& $\mathrm{HG}_{2,0}$ & $\mathrm{HG}_{2,2}$ \\
\hline
 Amplitude & 1 & $3.888\cdot 10^{-5}$ & $3.881\cdot 10^{-5}$ & $3.231\cdot 10^{-7}$ \\ \hline
 Phase [deg] & 0& -90.0 & 90.0 & -89.7  \\
 \hline
\hline
\end{tabular}
\label{tab-modedecomposition}
\end{table}

The entire hyperbolic paraboloidal phase map is applied to the Gaussian beam at the waist by representing the incident beam amplitude profile and the resulting phase map as discrete matrices. The modulated beam is numerically decomposed in the Hermite-Gaussian mode basis and the amplitude and phase of the most dominant HG mode contents are obtained through numerical overlap integration utilizing the orthogonality of the HG mode basis for describing coherent paraxial beams. The results are listed in Table~\ref{tab-modedecomposition}. With the hyperbolic paraboloidal phase profile applied to the Gaussian beam, the most dominant HG modes other than the input $\mathrm{HG}_{0,0}$ mode are the second order $\mathrm{HG}_{2,0}$ and $\mathrm{HG}_{0,2}$ modes. They have roughly the same amplitude, from the mode decomposition, but with a phase difference of $180^{\circ}$. The next dominant HG mode from the beam wavefront curvature modulation is the 4-th order $\mathrm{HG}_{2,2}$ mode, with an amplitude that is two orders of magnitude smaller than the amplitude of the second-order modes.

Due to the opposite phase of the $\mathrm{HG}_{2,0}$ and $\mathrm{HG}_{0,2}$ modes in the RF sidebands, the resulting RFL mode mismatch error signals derived from the second-order modes cancel out after detection and demodulation. Therefore, such an electro-optic lens device with alternating polarity electrodes on its own cannot provide the required second-order modes with the correct phase alignment for the RFL mode mismatch sensing. 

\begin{figure}[tbp]
    \centering
    \includegraphics[width=1\linewidth]{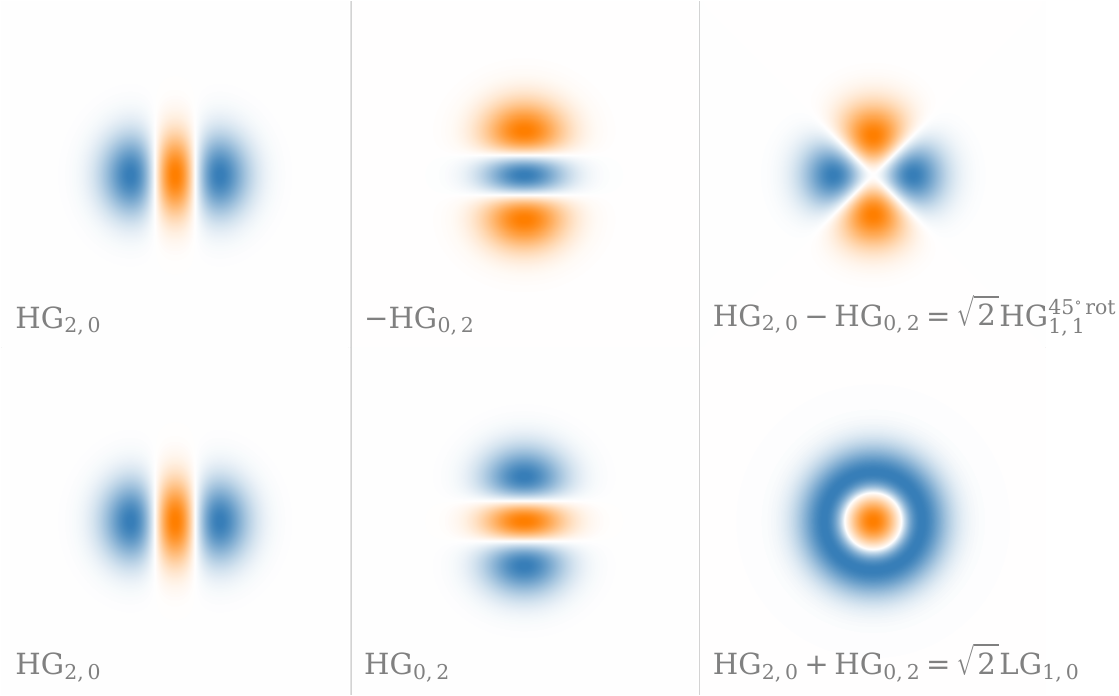}
    \caption{Mode converter made from cylindrical lenses rephases the second-order mode $\mathrm{HG}_{2,0} - \mathrm{HG}_{0,2}$ to $\mathrm{HG}_{2,0} + \mathrm{HG}_{0,2}$. This converts the RF $\mathrm{HG}_{1,1}^{45^\circ\, \text{rot}}$ mode to the desired $\mathrm{LG}_{1,0}$ mode for the RFL mode mismatch sensing.}
    \label{fig-mode_converter}
\end{figure}

Instead, one can solve this problem by mode matching the beam passing through the EOL into a subsequent $\pi/2$ mode converter telescope, as illustrated in Figure~\ref{fig-EOL_design}. The output beam through the EOL forms a new beam waist $w_{n}$ centered in the mode converter telescope, which consists of two astigmatic cylindrical lenses spaced by $\sqrt{2} \text{f}$ where f is the focal length and is related to the new waist size via $\text{f}=\frac{\pi w_{n}^2}{\lambda} / (1 + \frac{1}{\sqrt{2}})$~\cite{ONEIL200035, PhysRevD.100.102001}. As a result, it corrects the opposite phase of the pair of second-order modes generated by the EOL. Specifically, with an opposite phase in the $\mathrm{HG}_{2,0}$ and $\mathrm{HG}_{0,2}$ modes, the converted beam after the RF wavefront modulation is a 45-degree-rotated $\mathrm{HG}_{1,1}$ mode, as illustrated on the top panel of Figure~\ref{fig-mode_converter}. 
\begin{equation}
    \mathrm{HG}_{1,1}^{45^{\circ}\, \text{rot}} = \frac{1}{\sqrt{2}}\left(\mathrm{HG}_{2,0} - \mathrm{HG}_{0,2} \right)
\end{equation}
After the beam passes through the astigmatic mode converting telescope, it accumulates an extra phase of $\pi/2$ for the cylindrical lens focusing axis ($x$-axis) for each mode order in that direction compared to the normal Gouy phase accumulation for the non-focusing axis ($y$-axis)~\cite{PhysRevD.100.102001}. Since the second
order mode $\mathrm{HG}_{2,0}$ has two mode orders in the $x$-axis, it accumulate twice the extra Gouy phase $\pi/2$, i.e. $\pi$, compared to the $\mathrm{HG}_{0,2}$ mode. The sign of the $\mathrm{HG}_{2,0}$ mode as a result is flipped since $e^{i\pi}=-1$. The second-order modes $\mathrm{HG}_{2,0}$ and $\mathrm{HG}_{0,2}$ after the cylindrical lenses have the same phase
\begin{equation}
    \mathrm{LG}_{1,0} = \frac{1}{\sqrt{2}}\left(\mathrm{HG}_{2,0} + \mathrm{HG}_{0,2} \right)
\end{equation}
The second-order modes get rephased and converted to the desired $\mathrm{LG}_{1,0}$ mode for the RFL sensing, as shown in the bottom panel of Figure~\ref{fig-illustration}.

\section{RF Second-order Mode Generation}
\label{sec-2}
The second-order mode generation through the interaction of the Gaussian beam with the paraboloidal phase profile depends on the location of the interaction, due to the wavefront curvature and the beam size evolution of the input Gaussian beam. We would thus have to consider both the amplitude and phase of the second-order modes generated throughout the interaction with the electro-optic crystal region and add all the contributions to properly characterize the wavefront modulation efficiency of the EOL. 

\begin{figure}[tbp]
    \centering
    \includegraphics[width=1\linewidth]{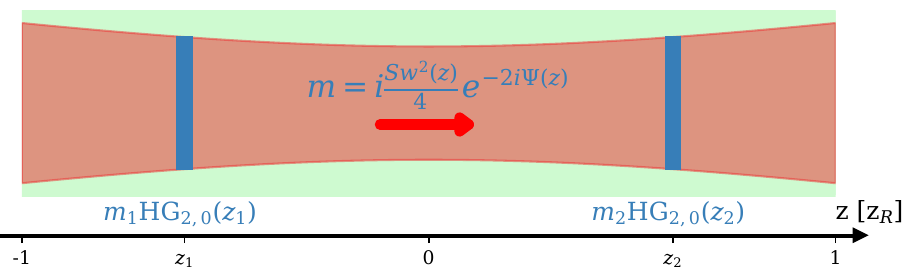}
    \caption{Illustration of the interaction between the input Gaussian beam in red with the EOL crystal in green. The amplitude and phase of the RF second-order mode coefficient $m(z)$ in general depends on the location of the interaction. }
    \label{fig-illustration}
\end{figure}

In Figure~\ref{fig-illustration}, the electro-optic crystal and the input Gaussian beam are illustrated in green and red respectively. For our design choice, the beam waist is placed at the center of the crystal. In general, in the HG mode basis, the field amplitude of second-order modes generated from the interaction with a crystal region around a given $z$ location can be characterized by a complex number $m(z)$: the mode scattering coefficient. The amplitude and the phase of the mode scatterng coefficient, in general, is a function of the $z$ location, due to the size and the wavefront curvature evolution of the input Gaussian beam. To obtain the total second-order mode content from the RF wavefront curvature modulation, we have to accumulate all the mode scattering coefficients throughout the interaction length within the crystal, by integrating the complex coefficients $m(z)$ over the entire crystal length.

As demonstrated in the previous section, the combination of EOL and mode converter effectively generates a wavefront curvature modulation to the input Gaussian beam. The curvature modulation can be characterized by applying the following phase factor to the Gaussian beam amplitude
\begin{equation}
e^{ik\frac{S}{2}(x^2+y^2)}
\label{equ-EOLphase}
\end{equation}
where the amount of curvature modulation is denoted as $S$. As derived in our previous work.~\cite{Tao:21}, for a generic HG mode with mode index $n$ in the $x$ direction, the effect of this extra phase factor on the $x$ component of the input beam amplitude can be expressed as mode scattering into the HG modes that are offset from the original mode by two mode orders
\begin{equation}
\begin{aligned} 
\mathcal{U}_{n}(x, z) & \approx \mathcal{U}_{n}+i k\frac{S w(z)^2}{8}\left(\sqrt{(n+1)(n+2)} \cdot \mathcal{U}_{n+2} e^{-2 i \Psi} \right. \\
&+(2 n+1) \cdot \mathcal{U}_{n} \left.+\sqrt{n(n-1)} \cdot \mathcal{U}_{n-2} e^{2 i \Psi}\right)
\end{aligned}
\end{equation}
We thus see the mode coefficients are $z$-dependent, in both the amplitude and the phase: the amplitude of the second-order modes scattered from the RF wavefront curvature modulation are proportional to the beam size squared, and the phases contain two factors of Gouy phase delay $\Psi$. 

In the special case of input $\mathrm{HG_{0,0}}$ mode, on which we focus in this paper, the second-order mode content after applying the phase factor can be written as
\begin{equation}
\begin{aligned} 
\text{2nd Order Modes} &= i k \frac{\sqrt{2} S  w(z)^2}{8}e^{-2 i \Psi} \cdot \left(\mathrm{HG}_{2, 0} + \mathrm{HG}_{0, 2}\right) \\
&= i k\frac{S w(z)^2}{4}e^{-2 i \Psi} \cdot \mathrm{LG}_{1,0}
\end{aligned} 
\end{equation}
in the HG or the LG mode basis. For instance, from the second-order mode amplitude in Table~\ref{tab-modedecomposition} with the approximation that the entire phase map is applied at the beam waist, we can calculate the effect of the phase map on the curvature modulation $S_{0}$ with $\pm 1$ V applied to the electrodes is
\begin{equation}
 S_{0} = \frac{\text{2nd Order Mode Content}}{k \sqrt{2} w_{m}^2/8} = 0.011 \, m^{-1}
\end{equation}
where we used $w_{m} = 58.3 \, \mu m$. Our EOL design of the three pairs of alternating polarity electrodes of $\pm 1$\,V effectively produces a wavefront curvature modulation of 11\,mD. 

The complex coefficient of the second-order modes generated from the interaction with the crystal of unit length around location $z$ thus is 
\begin{equation}
m(z) = i k \frac{S w(z)^2}{4}e^{-2 i \Psi} 
\label{equ-m(z)}
\end{equation}

Around the region [$z$, $z+dz$] in the crystal, the infinitesimal second-order mode generated is 
\begin{equation}
\begin{aligned} 
d m &= i k\frac{w(z)^2}{4}e^{-2 i \Psi} dS  \\
& = i k\frac{C w_0^2\left(1 + \left(\frac{z}{z_R}\right)^2\right) \cdot dz}{4}e^{-2 i \arctan{\frac{z}{z_{R}}}} 
\label{equ-dm2}
\end{aligned} 
\end{equation}
where we have substituted the $z$ dependence of the beam size and Gouy phase. We have also assumed a uniform electric field distribution along the beam propagation direction $z$ inside the EOL crystal so that the infinitesimal curvature modulation from crystal region [$z$, $z+dz$] can be written as $d S = C \cdot dz$, where the curvature modulation from unit crystal length $C$ is a constant. If we assume the total amount of beam wavefront curvature modulation generated from the entire EOL crystal is $S_0$, it then leads to
\begin{equation}
C = \frac{S_{0}}{L_z}
\end{equation}
assuming uniform electric field distribution, where $L_{z}$ is the crystal length.

Integrating along the region of the crystal [$-L_z/2$, $L_z/2$], we obtain the total amplitude of second-order modes produced: 
\begin{equation}
\begin{aligned}
m_{2} &= i k\frac{C w_0^2}{4} \int_{-\mathrm{L_z}/2}^{\mathrm{L_z}/2} \left(1 + \left(\frac{z}{z_R}\right)^2\right)  e^{-2 i \arctan{\frac{z}{z_{R}}}} dz \\
& = i k \frac{C w_0^2}{4} \int_{-\mathrm{L_z}/2}^{\mathrm{L_z}/2} \left(1 + \left(\frac{z}{z_R}\right)^2\right)  \cos\left({2 \arctan{\frac{z}{z_{R}}}}\right) dz \\
& = i k \frac{S_{0} w_0^2}{4} \left(1 - \frac{\mathrm{L_z}^2 \lambda^2}{12 \pi^2 w_{0}^4}\right)
\label{equ-EOLHG20_ana}
\end{aligned}
\end{equation}
where we have used the relation $C = \frac{S_{0}}{\mathrm{L_z}}$. $S_{0}$ is the total amount of beam wavefront curvature modulation through the entire electro-optic crystal (a property which is to first order independent of the beam traversing the crystal), $w_{0}$ is the waist size of the input Gaussian beam, and $L_{z}$ is the crystal length.

The second-order mode generation has also been investigated numerically by applying the paraboloidal phase maps to the propagating Gaussian beam in terms of two-dimensional arrays, as illustrated previously. For the numerical calculation, we divide the entire EOL crystal length along the beam propagation direction $z$ evenly into many segments and evenly distribute the paraboloidal phase map factor for the wavefront curvature modulation for each $z$ segment. The second-order mode amplitudes generated from each $z$ segment are then calculated and coherently summed over for the total second-order modes.

\begin{figure}[htbp]
    \centering
    \includegraphics[width=1\linewidth]{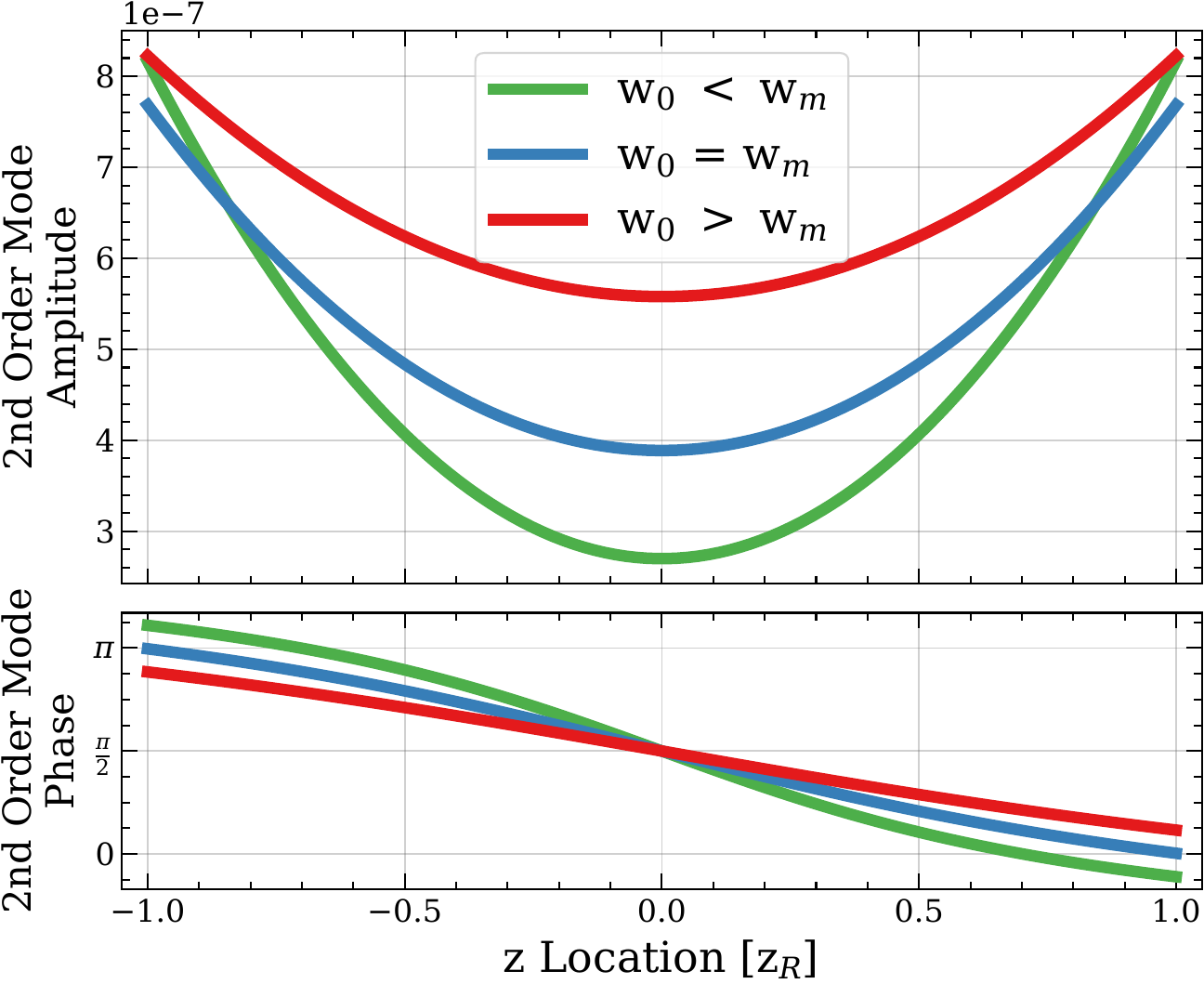}
    \caption{The amplitude and phase of the second-order mode coefficient as the entire crystal length is divided into 100 segments in the Gaussian beam propagation direction. The amplitude of the second-order mode is proportional to the square of the beam size at the interaction location. The phase difference between the second-order modes generated at different locations can lead to cancellation.}
    \label{fig-2ndampphase}
\end{figure}

Figure~\ref{fig-2ndampphase} shows the amplitude and the phase of the second-order mode content for all the $z$ segments throughout the intersection of the crystal, from the numerical calculation. The entire crystal range in $z$ was divided into 100 segments, which was shown to lead to a converging result in the total second-order mode generation. Three different cases corresponding to different waist sizes have been shown. The blue line represents the nominal waist size $w_{m} = 58.3 \, \upmu$m, and the red and green lines show the cases when the waist size is slightly larger and smaller than $w_{m}$ respectively. The amplitude of the second-order modes scales as the beam size squared. The phase of the second-order modes is depicted as 
\begin{equation}
\mathrm{arg\{m}_{2} \} = \frac{\pi}{2} - 2 \Psi(z)
\label{equ-HG20phase}
\end{equation}
as shown in Equation~\ref{equ-m(z)}, where $\pi/2$ comes from the prefactor $i$. For the case with the nominal waist size where the crystal length is exactly two Rayleigh lengths around the waist, as shown in the blue curve, the Gouy phase evolves from $-\pi/4$ at the front of the crystal, to 0 when at the center of the crystal, and to $\pi/4$ at the end of the crystal edge. The phase of the second-order mode is thus centered at $\pi/2$, from Equation~\ref{equ-HG20phase}. The range of the phase difference for the second-order modes generated at different locations is $\pi$, from twice the Gouy phase factor. On the other hand, for the large waist size solution in the red curve, the range of the phase difference is smaller than $\pi$, since the Gouy phase accumulation inside the crystal, in this case, is less than $\pi/2$. The phase range is larger than $\pi$ for the small waist size solution in the green.

From the amplitude and phase of the second-order modes generated from the interaction with each $z$ segment of the EOL crystal, we can then sum over all the contributions to get the total result. As can be seen, treating the entire phase map accumulation at a single waist location leads to inaccurate results. The inaccuracy comes from two factors: the amplitude of second-order mode generation depending on the beam size; and the phase mismatch between the second-order modes generated at different locations, which leads to some cancellation due to the phase difference. 

\section{Result and Discussion}
\label{sec-3}
The effects of beam size evolution and the phase mismatch cancellation on the total second-order mode generation can also be extracted from the analytical result in Equation~\ref{equ-EOLHG20_ana}. The first term $ik \frac{S_{0} w_0^2}{4}$ is the result if the entire interaction between the crystal and the input Gaussian beam is treated as a single interaction at the beam waist. Our more accurate result by treating the interaction throughout the crystal region individually differs from the simple treatment by the second term, which is a joint result of phase mismatch between the second-order modes generated at different locations along the longitudinal direction and the quadratic Gaussian beam size evolution throughout the crystal. 

\begin{figure}[htbp]
    \centering
    \includegraphics[width=1\linewidth]{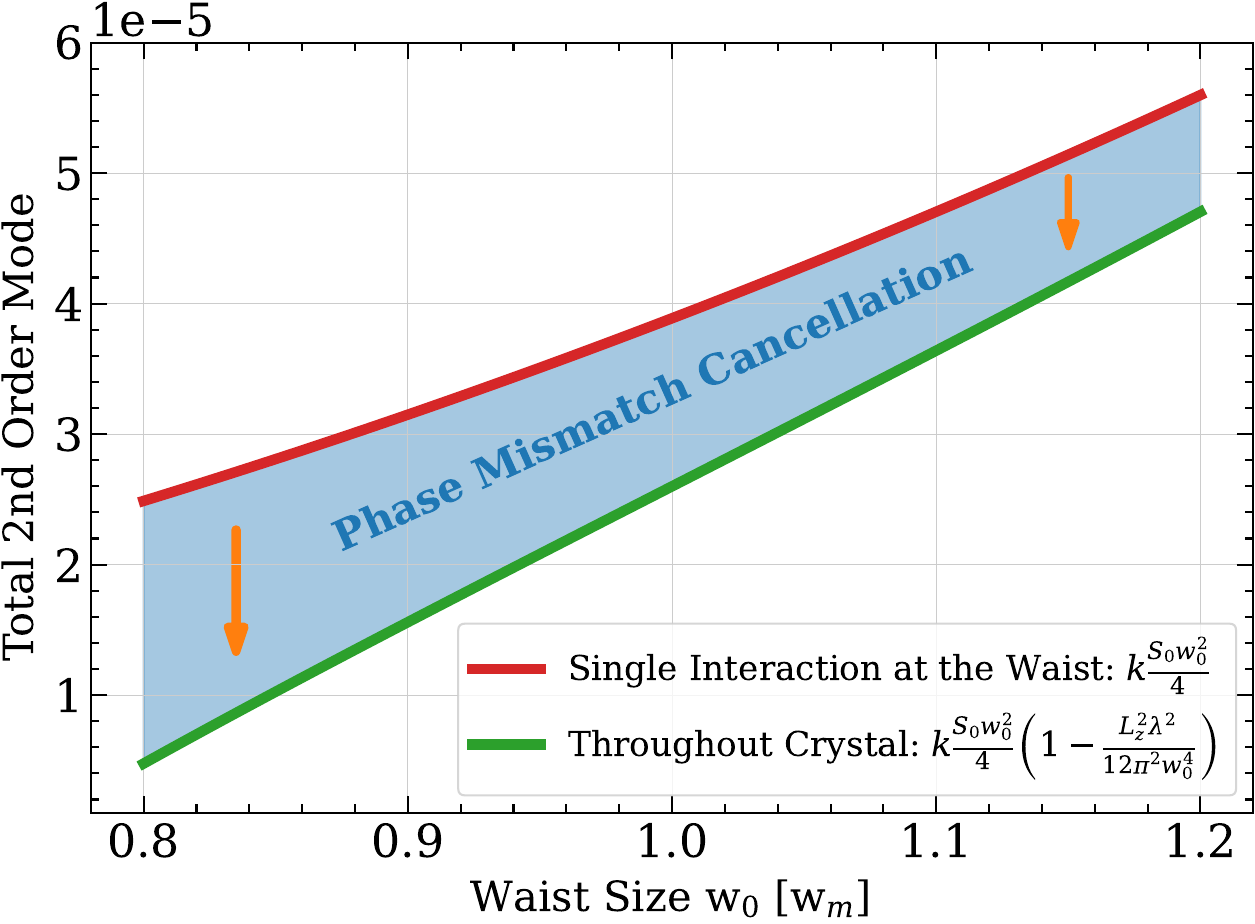}
    \caption{The total second-order mode content generated through the wavefront curvature modulation as the waist size increases. The interaction with the paraboloidal phase map is calculated in two approaches. The second-order mode calculated with the entire phase map interaction is treated at a single location at the beam waist is shown in red, which gives inaccurate results and leads to overestimation of the second-order mode generation compared to the more accurate approach with the interaction treated throughout the crystal, shown in green.}
    \label{fig-EOL_Phase_Mismatch}
\end{figure}

Figure~\ref{fig-EOL_Phase_Mismatch} shows the total second-order mode generation, from the two treatments. The resulting total second-order mode generation by treating the entire interaction of the input Gaussian beam with the phase map at the waist location is shown in the red curve, and the more precise calculation by treating the interaction throughout the crystal is shown in green. For instance, at the nominal waist size $w_{m}$, with the approximated treatment of a single interaction at the waist, the amount of second-order mode is roughly $3.8\cdot10^{-5}$, agreeing with the result from Table~\ref{tab-modedecomposition}. With a more careful and precise calculation by treating the interaction individually throughout the crystal, as shown in Figure~\ref{fig-2ndampphase}, the total second-order mode generation is roughly $2.5\cdot10^{-5}$, which is more than 35\% different from the approximated approach.

\begin{figure}[htbp]
    \centering
    \includegraphics[width=1\linewidth]{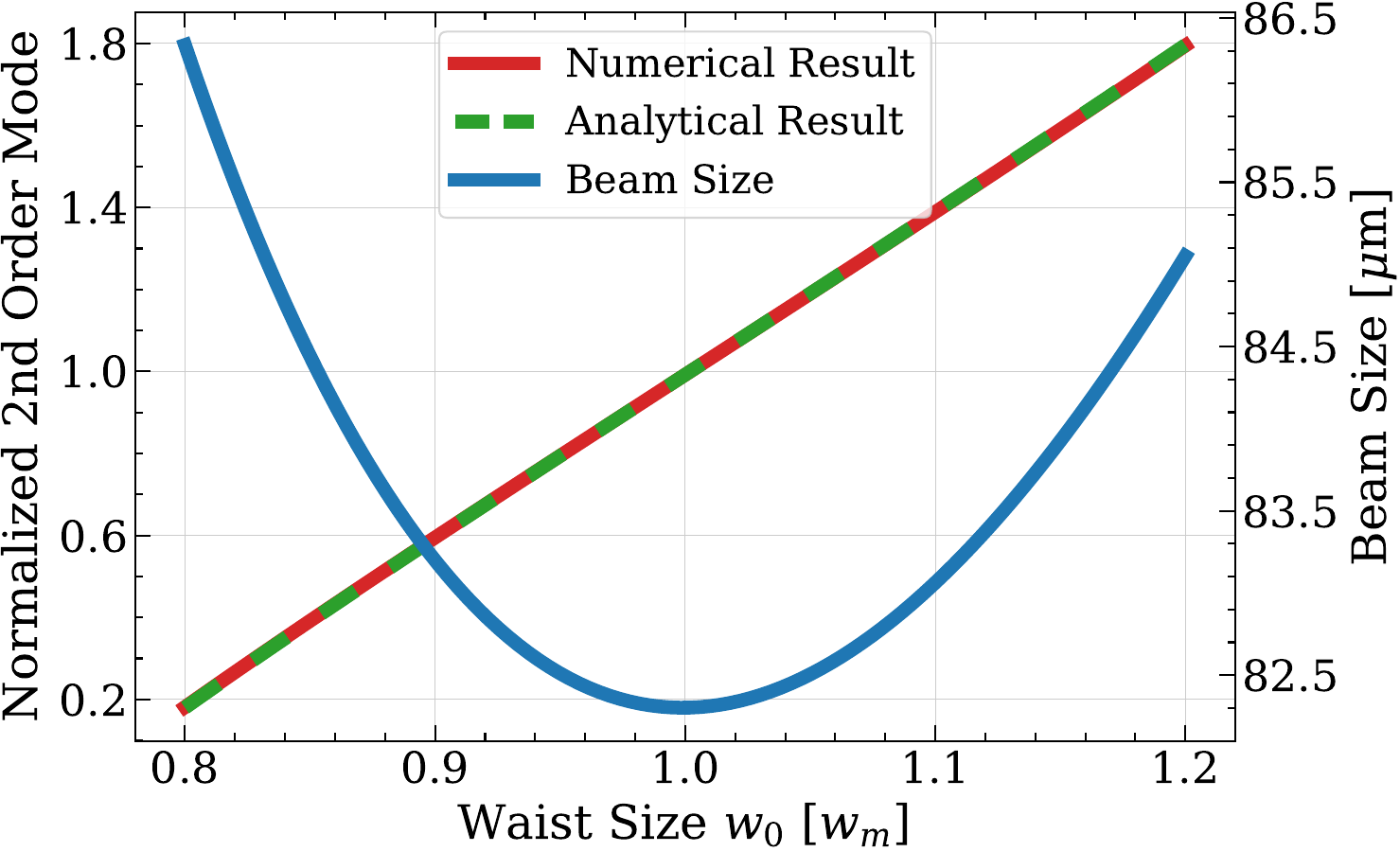}
    \caption{The total second-order mode generation, normalized by the case with the nominal waist size $w_{m}$, as a function of the waist size. The numerical results in red agree well with the analytical results in green. The corresponding beam size at the crystal edges is shown on the right y-axis.}
    \label{fig-EOL_waist_size_design}
\end{figure}

Figure~\ref{fig-EOL_waist_size_design} shows the normalized total amount of second-order mode amplitude, or the EOL wavefront curvature modulation efficiency, as a function of the waist size. The numerical result in the red curve agrees extremely well with the analytical result from Equation~\ref{equ-EOLHG20_ana} in the green dashed line. The second-order mode content for a given crystal length $L_{z}$ and total curvature modulation $S_{0}$ increases as the waist size increases, The merit of a larger waist size comes in two parts. Firstly the interaction with the phase profile generated from the EOL crystal is stronger with a larger average beam size throughout the crystal. In addition, the effect of mode cancellation from the phase mismatch between the second-order modes generated from different parts of the crystal is smaller due to smaller Gouy phase accumulation in the crystal range. 

The result for the second-order modes is normalized by the nominal waist size case with the smallest beam size at the crystal edges, which is shown on the right y-axis. For instance, if we increase the waist size from the optimal case by 10\%, the amount of second order modes, or the EOL wavefront curvature modulation efficiency, is increased by 40\%, at the expense of increasing the beam size at the crystal edges by roughly 1\,$\upmu$m, from the smallest beam size at 82.3\,$\upmu$m to roughly 83.3\,$\upmu$m. As a result, one can substantially increase the resulting RF beam wavefront modulation efficiency through a careful choice of the incident beam profile while maintaining a negligible increase in clipping losses.

\section{Conclusion}
\label{sec-4}
In this paper, we have proposed a realistic design and characterization of a novel Electro-Optic Lens device for high-efficiency generation of RF beam wavefront modulation. Upon inclusion of an additional mode converter telescope, it is able to generate the required second-order modes in the RF sideband fields with the correct relative phases and high efficiency. The second-order modes in the RF sideband fields can be used to beat with the second-order modes in the carrier field generated from static mode mismatch using single-element photodiodes to extract the full linearized mode mismatch sensing error signals in orthogonal demodulation phases, as demonstrated in the recently proposed QPD-free mode mismatch sensing scheme~\cite{master_thesis, PhysRevD.108.062001}. This scheme could be implemented to maintain optimal mode matching states and minimize optical loss in high-precision optical cavity experiments such as future interferometric gravitational wave detectors.

We demonstrated our approach to generating the required RF second-order HG modes for the RFL mode mismatch sensing with our design of an electro-optic lens followed by an astigmatic mode converter made from a pair of cylindrical lenses. We demonstrated our design for the EOL device by sandwiching an electro-optic crystal between three pairs of alternating polarity electrodes. The change in the index of refraction of the crystal calculated numerically with finite element methods was well characterized by a hyperbolic paraboloid function, which acts on the Gaussian beam passing through it by generating the second order $\mathrm{HG}_{2,0}$ mode and $\mathrm{HG}_{0,2}$ mode with roughly the same amplitude but the opposite phase. To rematch the phase of the second-order modes and generate a non-vanishing mode mismatch sensing signal, the EOL device is followed by a mode converter, which flips the sign of one of the second-order modes and realigns the phase. 

We provided a detailed analytical investigation of the total second-order mode generation from the paraboloidal phase profile by integrating the second-order mode complex coefficients throughout the crystal region. The calculation treats the interaction between the Gaussian beam and the phase front modulation throughout the crystal region individually, as the amplitude and phase of the second-order modes depend on the Gaussian beam propagation. The result is confirmed by a numerical approach that treats the beam amplitude profile and the phase profile as two-dimensional arrays. The effect of beam size evolution and the phase mismatch between the second-order modes generated at different locations was shown to lead to significant correction to the simplified and inaccurate approach by considering the entire phase map actuation at a single location at the beam waist.

The total second-order mode generation is related to the RF wavefront modulation efficiency and the strength of the RFL mode mismatch sensing signal. It was shown to be a monotonically increasing function of the waist size of the beam, for a given crystal length and total phase front curvature actuation. This gives us guidance in designing the beam profile to improve the total second-order mode generation, through a ``trade-off'' consideration with the beam size at the edges of the crystal and the resulting clipping loss. For instance, for a crystal length of 20 mm in the beam propagation direction, upon an increase in the waist size from the nominal waist size by 10\%, the total second-order mode generation is increased by 40\%, at the expense of increasing the beam size at the edges of the crystal by 1 $\mu m$, from the smallest beam size at 82.3 $\mu m$ to roughly 83.3 $\mu m$.

A thorough theoretical discussion of the RFL mode mismatch sensing scheme through RF modulating the beam wavefront curvature and a derivation of the sensing signal for an arbitrary Hermite-Gauss mode has been demonstrated in our previous work~\cite{PhysRevD.108.062001}. With the novel design of an electro-optic lens device introduced in this paper and the methods incorporated in characterizing the RF wavefront curvature modulation efficiency, there is work remaining to be done in the future for a realistic experimental demonstration and verification of such designs. This includes optimization of the RF beam wavefront modulation efficiency through the coupled beam profile and characterization of the quadratic paraboloidal phase profile in terms of the amplitude and the purity of the generated second-order modes in the RF sidebands respectively. In addition, the novel electro-optic lens design proposed in the current work opens a new research and development pathway toward realizing QPD-free beam wavefront modulation-based mode mismatch sensing schemes. Thus, with custom-built EOLs, the validity and performance of the RF lens mode mismatch sensing scheme based on fast wavefront curvature actuation provided by the proposed EOL device can be demonstrated with coupled cavity setups, by simultaneously extracting the full mode mismatch error signals in orthogonal demodulation phases from single-element RF detectors.

\section*{Acknowledgments}
This work was supported by National Science Foundation grants PHY-1806461 and PHY-2012021.

\section*{Disclosures}
The authors declare no conflicts of interest.

\section*{Data Availability}
Data underlying the results presented in this paper are not publicly available at this time but may be obtained from the authors upon reasonable request.

\bibliographystyle{apsrev4-2-trunc}
\bibliography{paper}

\end{document}